\documentclass[apex]{jjap3}
\usepackage{newtxtext, newtxmath}

\title{Heat flux sensing by anomalous Nernst effect in Fe-Al thin films on a flexible substrate}
\author{Weinan Zhou$^{1}$\thanks{E-mail: ZHOU.Weinan@nims.go.jp} and Yuya Sakuraba$^{1,2}$\thanks{E-mail: SAKURABA.Yuya@nims.go.jp}}
\inst{$^{1}$National Institute for Materials Science (NIMS), Sengen, Tsukuba 305-0047, Japan

$^{2}$PRESTO, Japan Science and Technology Agency, Saitama 332-0012, Japan}

\abst{We performed a numerical analysis of the material parameters required for realizing a heat flux sensor exploiting the anomalous Nernst effect (ANE).
The results showed the importance of high thermopower of ANE ($S_{\text{ANE}}$) and small saturation magnetization.
This motivated us to investigate the effect of Al substitution of Fe on ANE and found $S_{\text{ANE}} =$ 3.4 $\mu$V/K in Fe$_{81}$Al$_{19}$ because of the dominant intrinsic mechanism.
Using this material, we made a prototype ANE-based heat flux sensor on a thin flexible polyimide sheet and demonstrated accurate sensing with it.
This study gives important information for enhancing sensor sensitivity.}

\begin{document}
\maketitle

A heat flux sensor that enables a quick detection of the magnitude and direction of heat flow is expected to be a crucial component of a smart thermal management system.
However, commercially available heat flux sensors using the Seebeck effect (SE) have limitations hampering wider application.
SE-based heat flux sensors use a serially connected matrix of thermocouples on a solid substrate or thick flexible sheet as a support.\cite{SEsen1,SEsen2}
As a result, the sensor usually has a large thermal resistance that disturbs the innate heat flow.
Its flexibility is also limited by its mechanical fragility.
As well, although the sensitivity is proportional to the sensor's size, the complex structure makes it difficult to enlarge.
To overcome these limitations, a flexible heat flux sensor based on a combination of the spin-Seebeck effect (SSE)\cite{SSE1,SSE2} and the inverse spin Hall effect (ISHE)\cite{ISHE1} has been proposed.\cite{SSEsen1}
The electric field in this design is perpendicular to the direction of heat flow, in contrast to the parallel relationship in the SE design.
Hence, a simple bilayer consisting of a metallic layer and a magnetic layer on a flexible sheet can detect heat flow without any patterning.
This design has been demonstrated in a ferrite Ni$_{0.2}$Zn$_{0.3}$Fe$_{2.5}$O$_4$/Pt bilayer on a 25-$\mu$m-thick polyimide sheet.
However, the sensitivity was only 0.98 nV/(W$\cdot$m$^{-2}$), about four orders of magnitude smaller that of an SE-based heat flux sensor.
For practical use, the sensitivity should be improved.
One way to do so is to introduce a thermopile structure with laterally connected thermocouples consisting of two different metallic wires with positive and negative spin-Hall angles.\cite{SSEsen2}
The voltage is proportionally enlarged by elongating the total wire length.

An alternative approach for a heat flux sensor is to use the anomalous Nernst effect (ANE).
ANE is a thermoelectric phenomenon that occurs in conductive magnetic materials having a finite magnetization, such as magnetic metals\cite{ANE1,ANE2,ANE3,ANE4,ANE5,ANE6,Fe4N,FeGa2,ANE7,ANE8,TbFeCo,CMGt,ANE9,FeGa} and semiconductors,\cite{ANE10} as well as antiferromagnetic materials with a tiny magnetization due to their non-collinear spin structure.\cite{ANE11,ANE12}
The electric field induced by ANE ($\textbf{\textit{E}}_\text{ANE}$) can be expressed as\cite{ANE13}
\begin{equation}
\textbf{\textit{E}}_{\text{ANE}} = S_{\text{ANE}}\textbf{\textit{$\nabla$}}T \times \left(\frac{\textbf{\textit{M}}}{|\textbf{\textit{M}}|}\right), \label{eq1}
\end{equation}
where $S_{\text{ANE}}$ is the thermopower of ANE, \textbf{\textit{$\nabla$}}$T$ is the temperature gradient and \textbf{\textit{M}} is the magnetization.
Eq.~(\ref{eq1}) indicates that $\textbf{\textit{E}}_\text{ANE}$ is perpendicular to the heat flow, as in SSE-based heat flux sensors, so the thermopile structure can also be used to increase the sensitivity of an ANE-based one.
In addition, since ANE occurs in the bulk region of the magnetic material that generates a constant voltage ($V_{\text{ANE}}$) against a constant heat flux density ($J_{\text{Q}}$), the increase in sensor resistance due to elongating the wire, which increases Johnson noise, can be easily avoided by simply increasing the thickness of the magnetic material without losing $V_{\text{ANE}}$.
In this study, we performed a numerical simulation to elucidate the required material parameters for realizing a practical ANE-based heat flux sensor.
This analysis led us to investigate the properties of Fe$_{100-x}$Al$_{x}$ (Al at.\% $x$ = $0\sim32$) thin films and find a large $S_{\text{ANE}}$ originating from the intrinsic mechanism.
Finally, we evaluated the performance of a prototype ANE-based heat flux sensor using the developed Fe-Al thin films on thermally oxidized Si (Si/SiO$_{\text{x}}$) substrates and flexible polyimide sheets.

\begin{figure}
\centering
\includegraphics{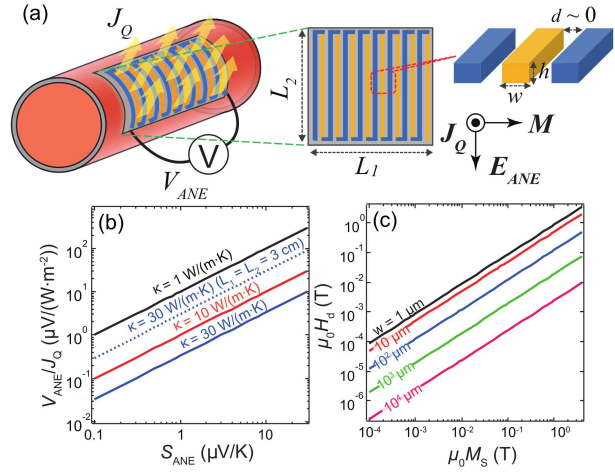}
\caption{(a) Schematic illustration of ANE-based heat flux sensor. (b) Calculated $S_\text{ANE}$ dependence of sensitivity $V_\text{ANE}/J_{\text{Q}}$ for different $\kappa$. (c) Calculated $M_\text{s}$ dependence of $H_\text{d}$ for different wire widths $w$.}
\label{Fig1}
\end{figure}

Figure~\ref{Fig1}(a) shows a schematic view of the ANE-based heat flux sensor having a thermopile structure.
To obtain a general expression for the sensitivity, we consider a rectangular-shaped sensing area ($=L_{\text{1}} \times L_{\text{2}}$) and laterally connected thermocouples consisting of wires of width $w$ and thickness $h$.
For simplicity, the neighboring wires $A$ and $B$ separated by a distance $d$ are made of the same material having $S_{\text{ANE}}$ and thermal conductivity $\kappa$.
Note that the thermocouple can be made of one material, because the direction of $\textbf{\textit{E}}_{\text{ANE}}$ can be controlled by the direction of $\textbf{\textit{M}}$.\cite{ANE4,ANE14}
The sensitivity of the sensor ($V_\text{ANE}/J_{\text{Q}}$) is proportional to $S_{\text{ANE}}$, the thermal resistivity ($= 1/\kappa$), and the sensing area:
\begin{equation}
\frac{V_{\text{ANE}}}{J_{\text{Q}}} = \frac{S_{\text{ANE}}L_{\text{1}}L_{\text{2}}}{\kappa(w+d)}. \label{eq2}
\end{equation}
To provide an indication of the parameters required for practical performance, we calculated the $S_{\text{ANE}}$ dependence of $V_\text{ANE}/J_{\text{Q}}$ for the case of $L_{\text{1}} = L_{\text{2}} =$ 1 cm, $w = h =$ 10 $\mu$m, and $d \sim$ 0 (negligibly small separation between the wires), as shown in Fig.~\ref{Fig1}(b).
The results indicate that for $\kappa =$ 10 or 30 W/(m$\cdot$K), $S_{\text{ANE}}$ over 10 $\mu$V/K or 30 $\mu$V/K is required to obtain a sensitivity comparable with a commercially available SE-based heat flux sensor having a 1 cm$^2$ sensing area ($\sim$ 10 $\mu$V/(W$\cdot$m$^{-2}$)\cite{sensor1}).
Although this value is higher than the highest $S_{\text{ANE}}$ reported so far ($\sim$ 6 $\mu$V/K\cite{ANE7,ANE9}), the sensitivity can be easily increased by expanding the sensing area (see the dotted line of $L_{\text{1}} = L_{\text{2}} =$ 3 cm in Fig.~\ref{Fig1}(b)).
Besides a high $S_{\text{ANE}}$, a small saturation magnetization ($M_\text{s}$) is also important for the ANE-based heat flux sensor.
To maximize $V_{\text{ANE}}$, $\textbf{\textit{M}}$ must be aligned to the width direction (Fig.~\ref{Fig1}(a)).
The magnetic anisotropy field of the material must withstand the demagnetization field ($H_\text{d}$) originating from $\textbf{\textit{M}}$ aligned to the width direction, which is not its magnetic easy axis.
Improving sensitivity by increasing the total wire length will result in narrower wires and lead to larger $H_\text{d}$ that may degrade the remanent magnetization along the width direction.
Fig.~\ref{Fig1}(c) shows the $M_\text{s}$ dependence of $H_\text{d}$ of a wire having a rectangular cross section of $h =$ 10 $\mu$m and $w =$ 1, 10, 10$^2$, 10$^3$, and 10$^4$ $\mu$m, which was calculated using an analytic expression.\cite{demag}
It shows that $M_\text{s}$ less than 0.2 or 0.8 T is needed to keep $H_\text{d}$ less than 0.1 T for $w =$ 10 or 10$^2$ $\mu$m.
Thus, a material showing a large ANE originating from its intrinsic Berry curvature that is beyond the positive scaling behavior of $S_\text{ANE}$ against $M_\text{s}$, e.g., Mn$_3$Sn\cite{ANE11} or Fe-Ga alloy\cite{FeGa}, would be a good choice.
Motivated by a recent report showing strong enhancement of ANE from pure Fe in Fe-Ga alloy, which is due to enlargement of the intrinsic contribution of ANE by Ga substitution,\cite{FeGa,FeCal} we investigated the effect of Al substitution of Fe on ANE and found a large $S_{\text{ANE}}$ in a material with the most abundant elements, which will be beneficial for a mass production.

The Fe-Al thin films were formed by codeposition of Fe and Al at room temperature using magnetron sputtering.
To investigate their properties, the Fe-Al thin films were deposited on MgO (100) single crystal substrates.
Samples of Fe-Al thin films on Si/SiO$_{\text{x}}$ substrates and 25-$\mu$m-thick flexible polyimide sheets were also prepared.
The composition of the films was determined by wavelength dispersive X-ray fluorescence analysis, while their structures were measured by X-ray diffraction (XRD).
The thickness of the thin films was obtained using the X-ray reflectivity method.
The magnetic properties were measured with a vibrating sample magnetometer.
For the anomalous Hall effect (AHE) and ANE measurements, the thin films were patterned into a Hall bar structure 3 mm wide and 7 mm long by photolithography and Ar ion milling.
The samples were set inside a physical property measurement system where the magnetic field ($H$) was applied in the out-of-plane direction of the substrate.
The ANE measurement used the method described in a previous paper\cite{FeGa} to get a reliable $S_{\text{ANE}}$ and $S_{\text{SE}}$.

\begin{figure}
\centering
\includegraphics{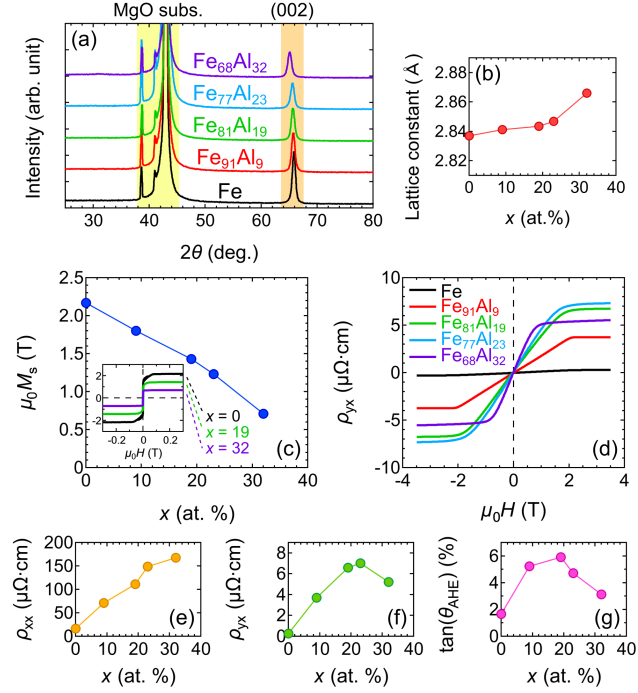}
\caption{(a) Out-of-plane XRD patterns of Fe$_{100-x}$Al$_{x}$ thin films. The signals within the yellow belt are from the MgO substrates, while the orange belt marks the (002) peak from the Fe$_{100-x}$Al$_{x}$ thin films. (b) Lattice constant of Fe$_{100-x}$Al$_{x}$ derived from (002) peak in (a) as a function of $x$. (c) $M_\text{s}$ of Fe$_{100-x}$Al$_{x}$ thin films on MgO (100) single crystal substrates as a function of $x$. The inset shows the $M$ - $H$ curves of Fe, Fe$_{81}$Al$_{19}$, and Fe$_{68}$Al$_{32}$ with $H$ applied in the film plane along the MgO [100] direction. (d) Anomalous Hall resistivity $\rho_\text{yx}$ as a function of $H$ along the out-of-plane direction. (e) Longitudinal resistivity $\rho_\text{xx}$ at zero $H$, (f) anomalous Hall resistivity $\rho_\text{yx}$, and (g) anomalous Hall angle tan($\theta_\text{AHE}$) = $\rho_\text{yx}$/$\rho_\text{xx}$ as a function of $x$.}
\label{Fig2}
\end{figure}

Fe$_{100-x}$Al$_{x}$ thin films with $x$ = 0, 9, 19, 23, and 32 were grown epitaxially on MgO (100) single crystal substrates, and the thickness were 34, 33, 35, 34, and 35 nm, respectively.
Figure~\ref{Fig2}(a) shows their out-of-plane XRD patterns.
Only the (002) peak from the simple body-centered cubic (bcc) structure was clearly observed.
The 2$\theta$ angle of the (002) peak was used to calculate the lattice constant, which showed a monotonic increase with increasing $x$ (Fig.~\ref{Fig2}(b)).
However, the change in the lattice constant was small, especially when $x \le$ 23.
The results indicate that random Al substitution of Fe did not cause any structural transformations in these sputtered thin films.
In contrast to the tiny structural variation, Al substitution significantly affected the magnetic and transport properties.
$M_\text{s}$ monotonically decreased with increasing Al substitution (Fig.~\ref{Fig2}(c)).
On the other hand, Al substitution dramatically enhanced both $\rho_\text{xx}$ and $\rho_\text{yx}$, as summarized in Figs.~\ref{Fig2}(d)-\ref{Fig2}(f).
The anomalous Hall angle tan($\theta_\text{AHE}$) = $\rho_\text{yx}$/$\rho_\text{xx}$ peaked at $x$ = 19 with a value of 5.9 \% (Fig.~\ref{Fig2}(g)).

\begin{figure}
\centering
\includegraphics{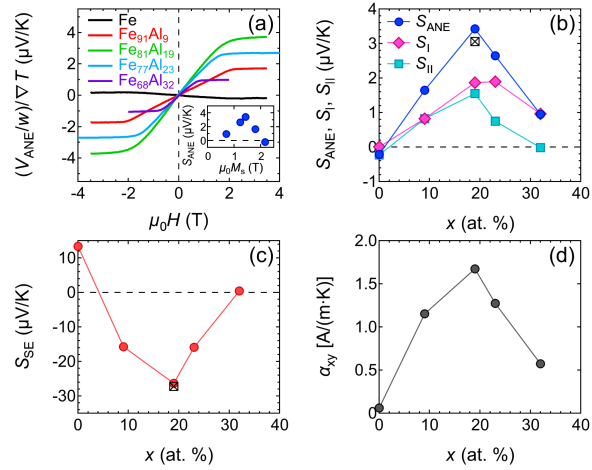}
\caption{(a) Normalized ANE voltage as a function of $H$ along the out-of-plane direction. The inset shows $S_\text{ANE}$ as a function of $M_\text{s}$. (b) $S_\text{ANE}$, $S_\text{I}$, $S_\text{II}$, (c) $S_\text{SE}$, and (d) $\alpha_\text{xy}$ as a function of $x$. The black crossed squares in (b) and (c) are from polycrystalline Fe$_{81}$Al$_{19}$ thin film on Si/SiO$_{\text{x}}$ substrate.}
\label{Fig3}
\end{figure}

Figure~\ref{Fig3}(a) shows the normalized ANE voltage from the Fe-Al thin films on MgO (100) single crystal substrates as a function of $H$ along the out-of-plane direction.
$S_\text{ANE}$ corresponds to the value of $(V_\text{ANE}/w)/$\textbf{\textit{$\nabla$}}$T$ linearly extrapolated to zero $H$ from the data points at high $H$, where $\textbf{\textit{M}}$ is saturated.
For the ANE voltage of Fe, an $H$-even component, which is attributable to the magneto-Seebeck effect, was subtracted.
Figures~\ref{Fig3}(b) and \ref{Fig3}(c) show the $x$ dependence of the $S_\text{ANE}$ and $S_\text{SE}$.
One can clearly see the significant enhancement of $S_\text{ANE}$ from pure Fe by Al substitution.
 $S_\text{ANE}$ reaches a maximum of 3.4 $\mu$V/K at $x =$ 19,, while $S_\text{SE}$ reaches a minimum.
The tendency of $S_\text{SE}$ as a function of $x$ is consistent with the previously reported results from bulk samples,\cite{FeAl} indicating the accuracy of the measurements obtained with our method.
$S_\text{ANE}$ can be separated into two components, $S_{\text{ANE}} = \rho_{\text{xx}}\alpha_{\text{xy}} - \rho_{\text{yx}}\alpha_{\text{xx}}$.\cite{ANE10,SANE}
The second term $- \rho_{\text{yx}}\alpha_{\text{xx}}$ (defined as $S_\text{II}$) originates from the AHE of the longitudinal carrier flow induced by SE and can be rewritten as $-S_{\text{SE}} \times \text{tan}(\theta_{\text{AHE}})$.
On the other hand, the first term $\rho_{\text{xx}}\alpha_{\text{xy}}$ (defined as $S_\text{I}$) is considered to be the intrinsic term of ANE, since the transverse Peltier coefficient $\alpha_{\text{xy}}$ directly converts \textbf{\textit{$\nabla$}}$T$ into a transverse current, $j_{\text{y}} = \alpha_{\text{xy}}\textbf{$\nabla$}T$.
Together with the results of the AHE measurement, we were able to derive $S_\text{I}$ and $S_\text{II}$ (Fig.~\ref{Fig3}(b)), as well as $\alpha_{\text{xy}}$ (Fig.~\ref{Fig3}(d)), which also peaks at $x =$ 19.
The $x$ dependence of $\alpha_{\text{xy}}$ can be partly explained as an enhancement of intrinsic $\alpha_{\text{xy}}$ due to Fermi-level shifting by Al substitution, as previously discussed in the case of Fe-Ga alloy.\cite{FeGa}
However, the maximum $\alpha_{\text{xy}}$ of the Fe-Al thin films is larger than that of Fe-Ga alloy, and currently the reason is not clear.
Meanwhile, due to both the larger $S_\text{II}$ from the larger tan($\theta_\text{AHE}$) and $-S_{\text{SE}}$, and the larger $S_\text{I}$ from the larger $\alpha_{\text{xy}}$, the maximum $S_{\text{ANE}}$ in the Fe-Al thin films is larger than that of the Fe-Ga alloy.
The inset of Fig.~\ref{Fig3}(a) plots $S_\text{ANE}$ as a function of $M_\text{s}$.
It shows that $S_\text{ANE}$ decreases with increasing $M_\text{s}$ when $\mu_{\text{0}}M_\text{s} >$ 1.4 T, which is similar behavior to that of Fe-Ga alloy\cite{FeGa} but different from the scaling behavior of various ferromagnetic materials,\cite{ANE11} suggesting that the dominant contribution to ANE is the intrinsic mechanism.
Since Fe$_{81}$Al$_{19}$ shows the largest $S_\text{ANE}$, this composition was used in the prototype ANE-based heat flux sensor.
To verify the properties of polycrystalline Fe$_{81}$Al$_{19}$ thin film, the same ANE measurements were carried out for the 35-nm-thick film prepared on Si/SiO$_{\text{x}}$ substrates.
The measured $S_\text{ANE}$ and $S_\text{SE}$, i.e., the black crossed squares in Figs.~\ref{Fig3}(b) and \ref{Fig3}(c), were not much different from their values for the epitaxial thin film.

\begin{figure}
\centering
\includegraphics{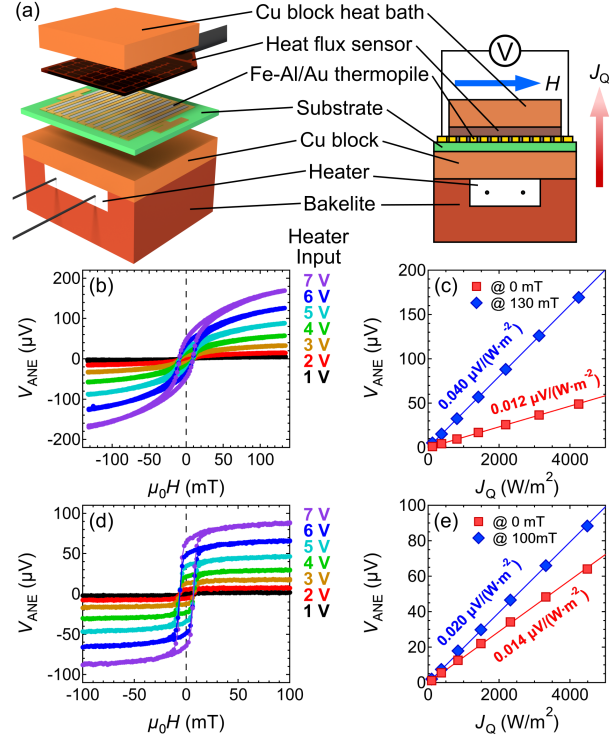}
\caption{(a) Schematic setup to evaluate the sensitivity of ANE-based heat flux sensor. (b) $V_{\text{ANE}}$ from the Fe-Al/Au thermopile on a Si/SiO$_{\text{x}}$ substrate as a function of $H$ for various heater inputs. (c) $V_{\text{ANE}}$ at $H =$ 0 and 130 mT as a function of $J_{\text{Q}}$. The solid lines are linear fittings through the origin. (d) $V_{\text{ANE}}$ from the Fe-Al/Au thermopile on a 25-$\mu$m-thick polyimide sheet as a function of $H$. (e) $V_{\text{ANE}}$ at $H =$ 0 and 100 mT as a function of $J_{\text{Q}}$.}
\label{Fig4}
\end{figure}

To make the ANE-based heat flux sensor, Fe-Al thin films on Si/SiO$_{\text{x}}$ substrates and 25-$\mu$m-thick polyimide sheets were patterned into parallel wires, followed by formation of Au electrodes to connect the wires in series.
The structure of the Fe-Al/Au thermopile was the same as the one in Fig.~\ref{Fig1}(a), except that only half the wires were made of magnetic material while the other half were for connection and had no ANE, forming so-called unileg thermocouples.
The setup to evaluate the sensitivity is shown in Fig.~\ref{Fig4}(a).
The ANE-based heat flux sensor and a commercially available heat flux sensor were sandwiched by two Cu blocks.
When a certain input voltage was applied to the heater, the bottom Cu block heated up while the top Cu block heat bath carried the heat away, creating a heat current in the out-of-plane direction.
After the setup reached thermal equilibrium, $H$ was swept in the thin film plane along the width direction of the wires, while the voltage from the ANE-based heat flux sensor was measured.
The measurements were carried out at room temperature.
Figures~\ref{Fig4}(b) and \ref{Fig4}(c) show the results for the Fe-Al/Au thermopile consisting of 25 Fe-Al wires on Si/SiO$_{\text{x}}$ substrate.
Each wire was 1-cm long and had $h =$ 500 nm, $w =$ 100 $\mu$m, and $d =$ 100 $\mu$m.
A clear hysteresis-like loop of the voltage due to the ANE of the Fe-Al wires was observed (Fig.~\ref{Fig4}(b)).
The magnitude of $V_{\text{ANE}}$ scaled with the heater input voltage.
$V_{\text{ANE}}$ at $H =$ 130 mT and in the remanent state (0 mT) is plotted in Fig.~\ref{Fig4}(c) as a function of $J_{\text{Q}}$ measured by the commercially available heat flux sensor.
The solid lines are linear fittings through the origin, which show a good linear relationship between $V_{\text{ANE}}$ and $J_{\text{Q}}$, demonstrating the feasibility of heat flux sensing using ANE.
The sensitivity of the ANE-based heat flux sensor was evaluated to be 0.040 $\mu$V/(W$\cdot$m$^{-2}$) at $H =$ 130 mT; however, it drastically decreased to 0.012 $\mu$V/(W$\cdot$m$^{-2}$) in the remanent state, which is attributable to misalignment of $\textbf{\textit{M}}$ from the width direction.
Figures~\ref{Fig4}(d) and \ref{Fig4}(e) show results for the Fe-Al/Au thermopile consisting of 10 Fe-Al wires on a 25-$\mu$m-thick polyimide sheet.
Each wire was 1-cm long and had $h =$ 300 nm, $w =$ 400 $\mu$m, and $d =$ 100 $\mu$m.
The Fe-Al wires were formed using a lift-off process, instead of Ar ion milling of a blanket Fe-Al thin film.
For an accurate measurement, the polyimide sheet was fixed to a Si/SiO$_{\text{x}}$ substrate, and a 0.5-mm-thick MgO substrate was placed on top of the Fe-Al/Au thermopile to ensure uniform heat current.
Hysteresis-like loops of $V_{\text{ANE}}$ were observed as a function of $H$ (Fig.~\ref{Fig4}(d)), but with better squareness than the ones from the sample on Si/SiO$_{\text{x}}$ substrate.
$V_{\text{ANE}}$ at $H =$ 100 mT and in the remanent state showed a good linear relationship with $J_{\text{Q}}$, as exhibited by the linear fittings through the origin in Fig.~\ref{Fig4}(e).
The sensitivity was 0.020 $\mu$V/(W$\cdot$m$^{-2}$) at $H =$ 100 mT, half that of the Fe-Al/Au thermopile on Si/SiO$_{\text{x}}$ substrate at $H =$ 130 mT, mainly due to the difference in the number of Fe-Al wires.
In the remanent state, the sensitivity decreased to 0.014 $\mu$V/(W$\cdot$m$^{-2}$), which is a much smaller decrease compared with Fig.~\ref{Fig4}(c).
This is partly due to the Fe-Al wires having a larger $w$, hence, a smaller $H_{\text{d}}$.
$\kappa$ of the material can be estimated using Eq.~(\ref{eq2}) and a known sensitivity, together with the $S_{\text{ANE}}$ and the geometry of the thermopile.
Using a sensitivity of 0.020 $\mu$V/(W$\cdot$m$^{-2}$) (Fig.~\ref{Fig4}(e)) and $S_{\text{ANE}} =$ 3.1 $\mu$V/K for polycrystalline Fe$_{81}$Al$_{19}$ (Fig.~\ref{Fig3}(b)), we estimated $\kappa$ to be $\sim$16 W/(m$\cdot$K), which is comparable to the bulk Fe-Al alloy value,\cite{FeAl} thus verifying the sensitivity measured with our setup.
The sensitivity of the ANE-based heat flux sensor was more than one order of magnitude higher than the SSE-based one,\cite{SSEsen1} although it was still two to three orders lower than the commercially available heat flux sensor having the same 1 cm$^2$ sensing area.
However, we can greatly increase the sensitivity from that of the prototype, e.g., by increasing the density of the wires (by reducing $w$ and $d$), replacing Au with a different magnetic material having  $S_{\text{ANE}}$ with opposite sign, increasing the area of the sensor, etc.
In addition, more suitable magnetic materials having larger $S_{\text{ANE}}$ and smaller $M_\text{s}$ may emerge and lead to an adequate sensitivity in the remanent state.
It is also worth mentioning that the prototype sensor on the flexible polyimide sheet had a very low thermal insulance, $\sim$10$^{-4}$ m$^{2}\cdot$K/W, i.e., one order of magnitude smaller than that of commercially available heat flux sensor, mostly due to the thin polyimide sheet.

In summary, our numerical analysis showed the importance of high $S_{\text{ANE}}$ and small $M_\text{s}$ in magnetic materials to realize a practical ANE-based heat flux sensor.
This suggests that the intrinsic mechanism arising from the Berry curvature is a key to enhancing ANE and overcoming the scaling law of $S_{\text{ANE}}$ against $M_\text{s}$.
We found that Al substitution largely enhances the intrinsic contribution of ANE in Fe due to Fermi level tuning and leads to a large $S_{\text{ANE}} =$ 3.4 $\mu$V/K in Fe$_{81}$Al$_{19}$.
Finally, we demonstrated heat flux sensing using a prototype ANE-based heat flux sensor made from Fe$_{81}$Al$_{19}$ thin films on a flexible substrate.
Although there remains much room to improve the sensitivity, this study presented important information for future development of ANE-based heat flux sensors.  

\acknowledgement
The authors thank M. Murata, A. Yamamoto, H. Fujita, H. Nakayama, R. Iguchi, and K. Uchida for their valuable discussions and N. Kojima and B. Masaoka for technical support.
This work was supported in part by PRESTO ``Scientific Innovation for Energy Harvesting Technology'' (Grant No. JPMJPR17R5) and NEDO.

\end{document}